\documentclass[aps,twocolumn,floats,floatfix,showpacs,superscriptaddress]{revtex4}
\usepackage{epsfig}
\begin{document}
\title{Generalized Network Growth: from Microscopic Strategies to the Real Internet Properties}
\author{Guido Caldarelli}
\affiliation{INFM UdR ROMA1 Dipartimento di Fisica, Universit\`a di Roma
``La Sapienza'' Piazzale A. Moro 2 00185 Roma, Italy}
\author{Paolo De Los Rios}  
\affiliation{Institut de Physique Th\'eorique, Universit\'e de Lausanne
CH-1015 Lausanne, Switzerland}
\author{Luciano Pietronero} 
\affiliation{INFM UdR ROMA1 Dipartimento di Fisica, Universit\`a di Roma
``La Sapienza'' Piazzale A. Moro 2 00185 Roma, Italy}
\date{\today} 
\begin{abstract} 
In this paper we present a generalized model for network growth
that links the microscopical agent strategies with the large scale 
behavior. This model is intended to reproduce 
the largest number of features of the Internet network at the 
Autonomous System (AS) level.
Our model of network grows by adding both new vertices and 
new edges between old vertices. In the latter case a 
``rewarding attachment'' takes place
mimicking the disassortative mixing between small routers to larger ones. 
We find a good agreement between experimental data and the model for the 
degree distribution, the betweenness distribution, the clustering 
coefficient and the correlation functions for the degrees.
\end{abstract} 
\maketitle

Networks or graphs are mathematical entities composed by sites (or vertices) 
connected by links (or edges)\cite{B85}. Due to the apparent 
simplicity of such definition, many attempt have been made in order to 
describe very different physical situations within this framework. 
More interestingly new unexpected properties 
(with respect to the traditional approach of Random Graphs due to 
P. Erd\H{o}s and A. R\'enyi \cite{ER61}), have been found in a variety 
of different systems.  
Internet\cite{FFF99,CMP00},   
WWW\cite{HA99,BAJ99,KRRT99}, social structures\cite{M67,N01,ASBS00} 
and even protein interactions\cite{SM02} display a self-similar distribution 
$P(k) \propto k^{-\gamma}$ for the degree $k$ (i.e. the number of links per vertex).   
From a theoretical point of view some of the active ingredients that 
determine such self-similarity in the statistical properties of the degree 
have already been found\cite{AB02,re3,BB01,CCDM02}. 
Interestingly, graph properties are far more complex than the degree 
distribution. 
Therefore the onset of a complete set of topological properties in 
the above real networks remains to be fully explained. 
The most interesting case is represented by Internet, where a full 
understanding of the statistical properties of the phenomenon could 
help in improving the technical features. 

In this Letter we want to present a simple statistical model 
able to reproduce many statistical properties of the Internet 
beyond the degree distribution (most of the present models 
focus mainly on that). This is achieved, as explained 
below, by relating the microscopic agents strategies and the 
macroscopic statistical properties.

A way to describe the complex phenomenology (even by restricting 
to the topology, ignoring the different weight of various links 
given by the traffic) can be made by considering the clustering 
correlation and centrality present in the graph. 

{\em Clustering} measures the presence of part of the graph  
denser than the average. 
The most immediate measure of clustering is the clustering coefficient  
$c_i$ for every site $i$. This quantity 
gives the probability that two nearest neighbors of a vertex are also
neighbors to each other. 
The clustering coefficient can then be averaged over the vertices 
in the structure, giving the total clustering $<C>$ or rather be 
decomposed by considering the function 
$c(k)$ giving the probability that a vertex has clustering
coefficient $c(k)$  given its degree $k$. 

{\em Correlation} is best represented by the conditional probability 
$P_C(k'|k)$ that a link belonging to a node with degree $k$ points to a node
with degree $k'$. If this is independent on $k$, we have $P_C(k'|k)=P_C(k')
\simeq k'P(k')$. If instead there is a dependence on $k$ we can establish 
the strength of correlation between vertices of different degree. 
The most immediate way to compute such a correlation is given by considering 
the quantity 
\begin{equation} 
<k_{nn}> = \sum_{k'} k'P(k'|k) 
\end{equation} 
i.e. the nearest neighbors average degree of nodes with degree $k$.

{\em Centrality} of some vertices with respect to other zones 
is also a way  to consider deviations from average behavior in the structure. 
In particular betweenness and closeness, are the measures of the centrality of
a site with respect to the other vertices in the graph~\cite{Fre}.
The betweenness $b$ of a vertex $i$ gives the probability that the 
site $i$ is in the shortest path from vertex $j$ and vertex $k$ 
(for every $j$ and $k$). 
If the number of shortest paths between a pair of vertices $(j,k)$
is $D(j,k)$, we denote with $D_i(j,k)$ the number of such shortest
paths running through $i$.
The fraction $g_i(j,k)=D_i(j,k)/D(j,k)$ may be interpreted as
the amount of the role played by the vertex $i$ in social relation between
two persons $j$ and $k$. 
The betweenness of $i$ is defined as the sum of $g_i(j,k)$ over all 
the connected pairs. 
\begin{equation}
g_i=\sum_{j\neq k}g_i(j,k)=\sum_{j\neq k}\frac{D_i(j,k)}{D(j,k)}.
\end{equation}
  
All the above quantities have already been considered and analyzed in a 
series of papers \cite{Ale2,Ale}. 
The main result of these analysis is the detection of a complex inner 
structure resulting in a clustering larger 
than expected, a power law distribution of the $c(k)$ (i.e. $c(k) \propto 
k^{-\phi}$) a power law distribution of the correlation $k_{nn}(k)$ (i.e. 
$k_{nn}(k) \propto k^{-\tau}$) and finally a power law distribution of the 
values of the betweenness $b$ (i.e.  $P(b) \propto b^{-\eta}$ ). 
On the basis of these analysis the Internet at the level of Autonomous Systems (AS) 
shows a self-similar behavior in all these quantities signaling a possible 
presence of a critical state.
Interestingly, all the statistical models introduced do not reproduce 
such features.
The Barab\'asi Albert model introduced the concept of preferential attachment
such that the network grows by addition of new nodes that link with the 
older ones with a probability proportional to the degree of the latter ones.
Even if this simple rule reproduces nicely the degree distribution it 
fails in reproducing the correlations, the centrality and the clustering 
of the AS systems. 
Some modifications of the above model give a better qualitative 
agreement with the real data. In particular in the stochastic growth model 
proposed in \cite{g2} a constant fraction of site are added  at every 
timestep and a substantial rewiring takes place. 
Also the fitness model has a fairly nice agreement 
with the shape of the AS distributions. In this model\cite{gin} the  
preferential attachment is weighted through an individual site fitness.

The above models, anyway, do not give a precise quantitative prediction 
of all the properties measured in teh AS network. 
Here we want to present a new statistical model giving a better agreement
with the data and linking the microscopic dynamic to the macroscopic evolution. 
The basic idea of the model is to allow both the addition of a vertex 
(with probability $p$) and the addition of a link (with probability $1-p$). 
Typically 
such link relates two sites $1,2$ whose degrees are $k_1$ and $k_2$.
It is natural to define a "directionality" in the link. This is defined by 
deciding who pays the cost of the connections. 
This should mimic for example users that 
pay to get wired to Internet, flow of information etc. In this paper we will 
not consider directionality explicitly, leaving the extension to the oriented 
graphs to future work. 
In general, we can write the probability of addition of this link as 
$\tilde{P}(k_1,k_2)$. 
The specific form can be directly linked to the microscopical agents strategies.
There are two obvious limiting cases. The case $k_1=0$ corresponds to a 
new site which decides to join the network. The case $k_2=0$ 
corresponds to the creation of 
a link toward a site not connected to the network. 
The process is asymmetric due to the growth rules, that allows to write
$\tilde{P}(k_1,k_2)$ in terms of conditional probabilities. If
site $1$ pays for connection, then 
$\tilde{P}(k_1,k_2)=\tilde{P_2}(k_2|k_1) \tilde{P_1}(k_1)$. 
A simple ansatz corresponding to the BA model, would be to assume that 
$\tilde{P_1}(k_1)=\delta_{k_1,0}$ and $\tilde{P}(k_2|k_1) \propto k_2$. 
The generalization presented in 
Ref.\cite{re1,re2} 
assumes instead $\tilde{P_2}(k_2|k_1)=k_2+\lambda$ ($k_2$ is the in-degree) and 
$\tilde{P_1}(k_1)=k_1+\mu$ ($k_1$ is the out-degree). 

Here we propose to consider non oriented graphs (as it is the case of AS). 
Furthermore we assume $\tilde{P_1}(k_1)\propto k_1$ and we tune the form 
of the $\tilde{P_2}(k_1|k_2)$ in order to obtain the different 
situations observed in the experimental data\cite{Ale,Ale2}. 
For example a form of the type 
\begin{equation}
\tilde{P_2}(k_2|k_1)\propto \frac {1}{|k_1-k_2|+1}
\end{equation}
would produce the so-called assortative mixing\cite{N02} where vertices of the same degree 
tend to be connected to each other.

We instead focus on the opposite limit where  
\begin{equation}
\tilde{P_2}(k_2|k_1)\propto |k_1-k_2|
\end{equation}
in order to model the so-called disassortative mixing\cite{N02} with large hubs and 
poorly connected vertices.
In this limit at any time step
\begin{enumerate}
\item{Either a vertex is added and linked with vertex $i$ with probability}
\begin{equation}
p\frac{k_i}{\sum_{j=1,N} k_j}.
\end{equation}
\item{or an edge is added (if absent) between vertices $i$ and $j$
already present. with probability}
\begin{equation}
(1-p)\frac{k_i}{\sum_{k=1,N} k_k}\frac{|k_i-k_j|}{\sum_{k\neq i=1,N} |k_i-k_k|}.
\end{equation}
\end{enumerate}

From the above rules, the case $p=1$ (no edge 
creation) corresponds to a traditional AB model where only one edge is added 
for time step. 
Intuitively as the parameter $p$ is tuned to $0$ the edge growth becomes more and more important.
This results in a larger connected core with respect to
the AB model, as shown in Fig.1.

Numerical simulations of this model are presented below for different  
cases.  
The quantities we decide to monitor are the 
distribution of the degree $P(k)$, the distribution of both betweenness $b$ and
closeness $c$, 
the clustering coefficient $c(k)$ of vertices whose degree is $k$ and finally 
the average degree $,k_{nn}(k)$ of neighbours of a vertex whose degree is $k$.
All these quantities together with experimental data for the AS system
are reported in Table 1.

As reported in Fig.2, the probability distribution of the degree, 
$P(k)$, follows a 
power law behavior for every value of the parameter $p$.  

In particular the exponent $\gamma$ diminishes as loops start to form
in the system when $p < 1$.
Using such numerical evidence about the shape of the $P(k)$ we can present 
an analytical estimate of the exponent $\gamma$ through simple
arguments. We firstly notice that the number of edges in the system 
increases by one unity at every time step. Consequently the total degree 
over the network increases by two
\begin{equation}
\sum_{k=1}^{k_{max}} k N_k(t) = 2t.
\label{total degree}
\end{equation}
Here $N_k(t)$ is the number of vertices with degree $k$ at time $t$.

The total number of vertices, instead, increases at a rate $p$. 
Therefore the total number of vertices is 
\begin{equation}
\sum_{k=1}^{k_{max}} N_k(t) = pt
\label{total vertices}
\end{equation}
We assume that there is a stationary state, so that the number of vertices
grows linearly in time, $N_k(t) = n_k t$. 
As stated above,  we also assume that the degree distribution is a power law, 
$n_k = a k^{-\gamma}$, as seen from simulations in agreement with the 
preferential attachment rule. Then we can write
\begin{equation}
\sum_{k=1}^{k_{max}} n_k \simeq a\int_1^\infty k^{-\gamma} dk = \frac{a}{\gamma -1} = p
\label{prima eq}
\end{equation}
and
\begin{equation}
\sum_{k=1}^{k_{max}} k n_k \simeq a \int_1^\infty k^{-\gamma+1} dk = \frac{a}{\gamma -2} = 2
\label{seconda eq}
\end{equation}
Using Eqs.\ref{prima eq} and \ref{seconda eq} we obtain
$\gamma(p) = 2+\frac{p}{2-p}$, which provides good estimates for the results 
of the simulations, and correctly recovers the limiting cases, $\gamma(1) = 3$. 
A striking feature of the model is that as soon as $p$ is different form $0$ 
one still deals with a scale free network. 
The limit value for the distribution is $\gamma(0^+)=2$. 
The case $p=0$ is degenerate giving rise to a complete graph whose degree 
distribution is a delta function peaked around the size $n$ of the system. 
One can argue that this limit is peculiar. Indeed a complete graph of $N$ nodes
is characterized by a number of edges of order $N^2$.  

In our model, for arbitrary small but strictly positive $p$, both the
number of nodes and the number of edges grow linearly in time, with
a fixed ratio, so that the graph will never be complete. 
As regards the distance distribution, we find the small world effect, that is 

a peak around a characteristic value\cite{WS98}.

It has recently been shown\cite{Goh} that the probability  distribution 
$P(b)$ for the betweenness $b$ follows a power law,  
\begin{equation}
P_B(g) \sim g^{-\eta}, 
\end{equation} 
where $\eta=2$ is about $2$. 
For our model we find, in agreement with Ref. \cite{Goh}, that
the exponent $\eta$ is equal to $2.0$ if $p=1$. From the data for $p \ne 1.0$ 
we can conclude that the exponent changes to $\eta =2.2$,  
as it happens for the BA model when $m>1$ and loops start to appear in the
network (see inset of Fig.2). 
Another important measure of centrality is given by the closeness $c$. 
Closeness of a site $i$ is simply the inverse of the sum of the distances
from $i$ toward all the other vertices. Not surprisingly since the distance
distribution has a small-world effect this quantity has a frequency
distribution decreasing exponentially.

It is interesting to study the structure of such quantity with respect to 
the degree distribution. In particular we checked the behavior of $c(k)$ 
defined as the average clustering coefficient for a site 
whose degree is $k$ . Also this quantity could be fitted with a power 
law $c(k) \simeq k^{-\phi}$ as shown in Fig.3.
The model for $p=1.0$ is a BA tree and therefore by definition (since no loops 
are present) the clustering coefficient is always zero.
Instead in the BA model where $m$ is larger than 1, loops are present and 
the distribution of $c(k)$ with respect to $k$ is flat.
A very similar behavior can be found for the $<k_{nn}(k)>$ 
(inset of Fig.\ref{fig4}).
Again we have a power law of the form $<k_{nn}(k)>\simeq k^{-\tau}$ for 
large probability of
rewiring ($p<<1$) while this structuring disappears when 
$p=1$\cite{Ale,Ale2,mey}).

In conclusion we presented a model whose topological
features depend on the parameter $p$ tuning the ratio of vertices to 
edges creation.  Interestingly, we find that for
$p = 0.5(1)$ the model nicely reproduces  most of the properties 
measured in the real case. 
It would be then very tempting to assume that substantial rewiring 
in existing routers is the key ingredient that makes the statistical 
properties of Internet networks so different from other growing networks.

We thank the FET Open project IST-2001-33555 COSIN, the OFES-Bern under contract 
02.0234 and the Herbette Foundation for partial support.

\begin{figure} 
\centerline{\psfig{file=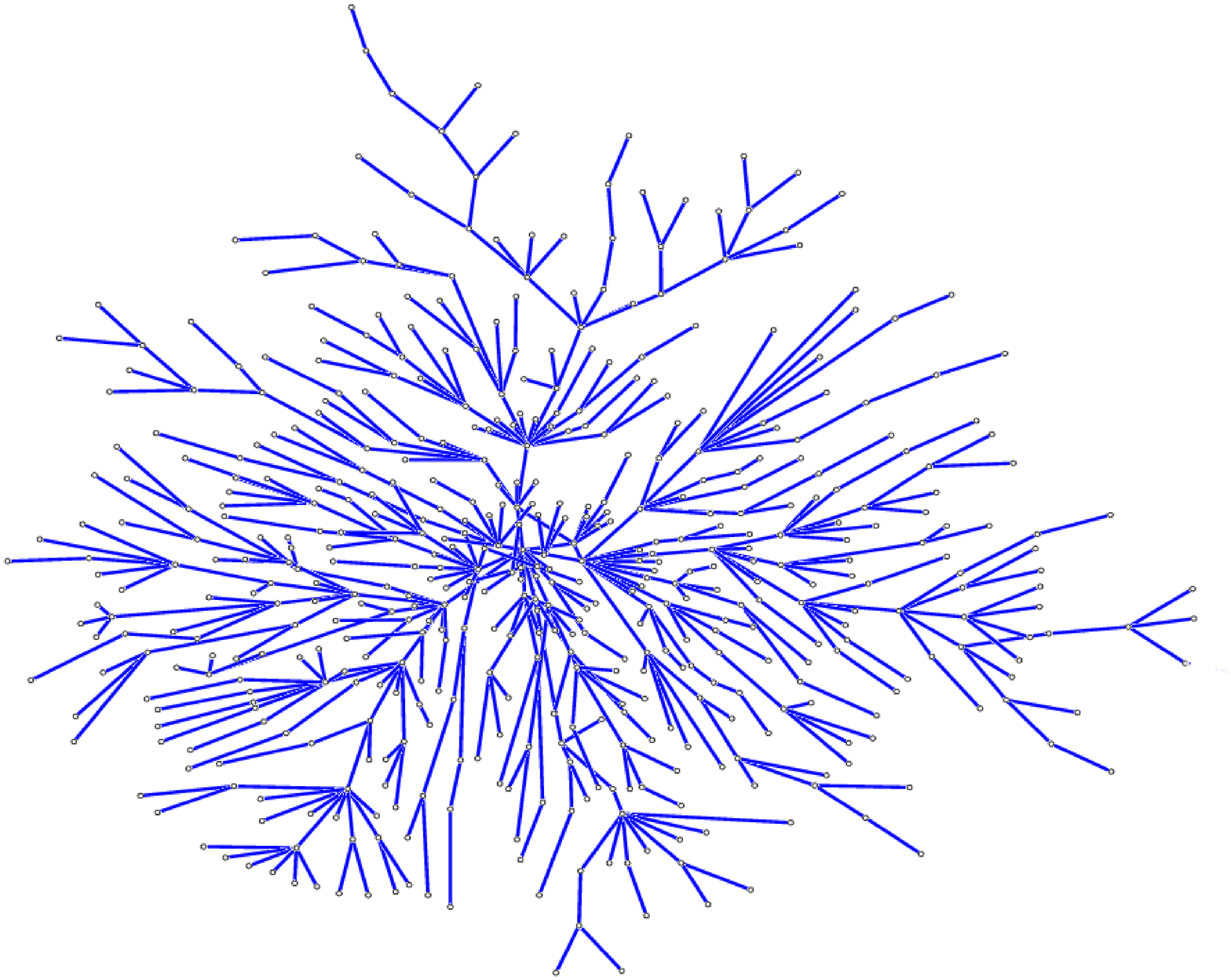,width=8cm,height=4.5cm}}
\centerline{\psfig{file=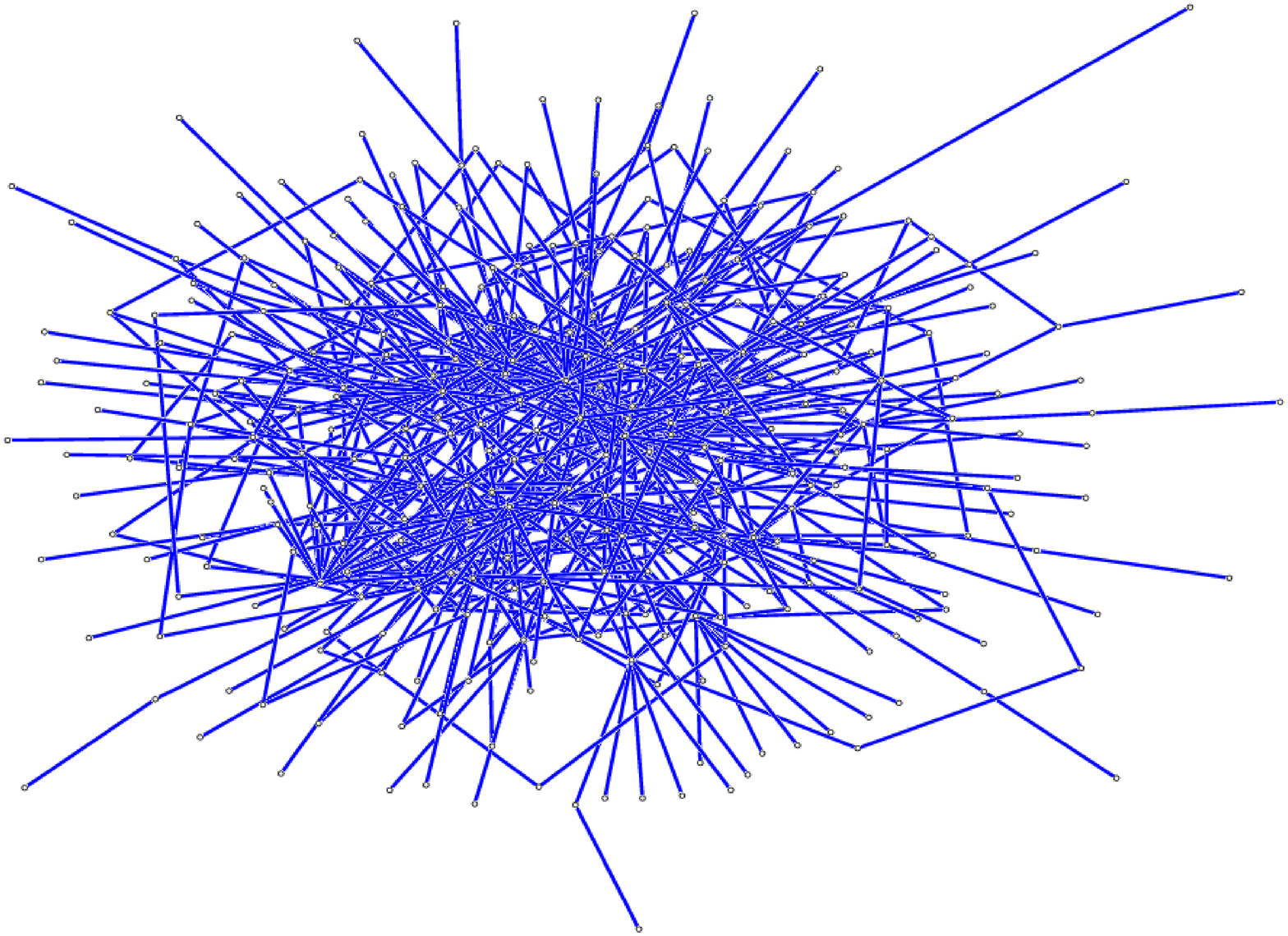,width=8cm,height=4.5cm}}
\caption{Plot of graphs obtained for different values of $p$. Above
a graph with $p=1$ corresponding to AB tree; below the rewiring 
produced by a $p=0.5$ simulation, gives rise to a more connected structure.
Pictures made with Pajek} 
\label{fig2} 
\end{figure} 
\begin{figure} 
\centerline{\psfig{file=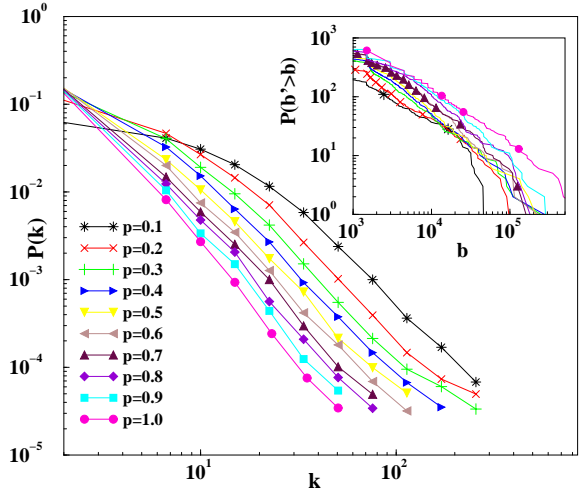,width=7.5cm}} 
\caption{Plot of the degree distribution for various values of $p$.
In the inset the integrated betweenness distribution. For the latter one only the symbols 
for $p=0.1$, $p=0.7$, $p=0.9$, $p=1.0$ have been explicitely plotted.} 
\label{fig3} 
\end{figure} 
\begin{figure} 
\centerline{\psfig{file=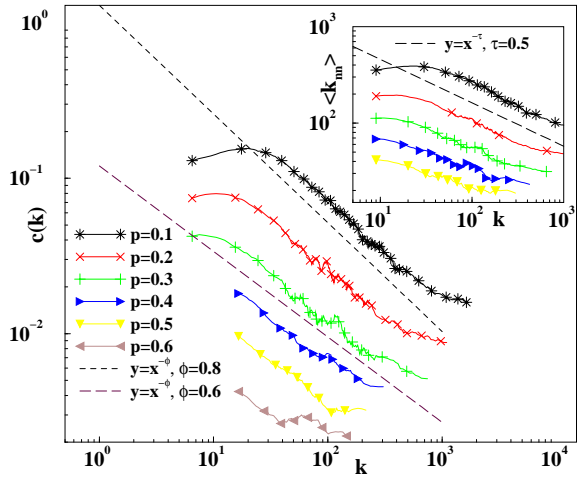,width=7.5cm}} 
\caption{Plot of the average clustering coefficient
for vertices whose degree is $k$. In the inset the average degree of the nearest
neighbours of a
vertex whose degree is $k$.}
\label{fig4} 
\end{figure} 
\begin{table} 
\begin{centering}
\caption{Data from Numerical simulation of the model. The last row refers to the
Internet (AS) network.}
\label{tab}
\begin{tabular}{|c|c|c|c|c|c|c|c|}
$p$   & $<k>$     & $\gamma$ & $2+\frac{p}{2-p}$& $<d>$     & $\eta$   & $\phi$ &$\tau$ \\
\hline
$0.1$ & $19.9(5)$ &$2.15(5)$ & $2.05  $         & $2.8(3)$  & $2.1(1)$ & $0.8(1)$& $0.5(1)$\\
$0.2$ & $10.0(3)$ &$2.2(1)$  & $2.11  $         & $2.9(3)$  & $2.1(1)$ & $0.8(2)$& $0.5(1)$\\
$0.3$ & $6.6(3)$  &$2.3(2)$  & $2.18  $         & $3.0(2)$  & $2.1(2)$ & $0.7(2)$& $0.5(1)$\\
$0.4$ & $5.0(2)$  &$2.3(3)$  & $2.25  $         & $3.1(2)$  & $2.2(1)$ & $0.7(2)$& $0.5(2)$\\
$0.5$ & $4.0(2)$  &$2.5(2)$  & $2.33  $         & $3.4(2)$  & $2.2(1)$ & $0.6(3)$& $0.5(2)$\\
$0.6$ & $3.3(2)$  &$2.5(2)$  & $2.43  $         & $3.9(3)$  & $2.1(1)$ & $0.5(4)$& $0.5(3)$\\
$0.7$ & $2.8(1)$  &$2.6(1)$  & $2.54  $         & $4.4(3)$  & $2.3(2)$ & $  -   $& $  -   $\\
$0.8$ & $2.5(1)$  &$2.7(1)$  & $2.67  $         & $5.5(1)$  & $2.1(1)$ & $  -   $& $  -   $\\
$0.9$ & $2.2(1)$  &$2.9(1)$  & $2.82  $         & $6.5(4)$  & $2.2(2)$ & $  -   $& $  -   $\\
$1.0$ & $2.0(1)$  &$3.0(1)$  & $3.00  $         & $8.7(3)$  & $2.0(1)$ & $  -   $& $  -   $\\
\hline
$AS$ & $3.8(1)$   &$2.22(1)$ & $ ---  $         & $4.16(1)$ & $2.2(1)$ & $0.75$&$0.55$ \\
\end{tabular}
\end{centering}
\end{table} 


\begin{thebibliography}{99} 
\bibitem{B85}  
B. Bollob\'as,  
{\it Random Graphs} (Ac. Press, London) (1985).  
 
\bibitem{ER61}  
 P. Erd\H{o}s, A. R\'enyi,  
{\em Bull. Inst. Int. Stat.} {\bf 38}, 343 (1961).  

  
\bibitem{FFF99}  
M. Faloutsos, P. Faloutsos, C. Faloutsos, 
 {\em Proc. ACM}, SIGCOMM (1999).  
 
\bibitem{CMP00} 
G. Caldarelli, R. Marchetti, L. Pietronero,  
 {\em Europhys. Lett.} {\bf 52}, 386 (2000).  
 
\bibitem{HA99}  
B.A. Huberman and L.A. Adamic 
{\em Nature} {\bf 399}, 130 (1999). 

\bibitem{BAJ99}  
A.-L. Barab\'asi, R. Albert and H. Jeong,  
{\em Physica A} {\bf 272}, 173 (1999).  
 
\bibitem{KRRT99}  
R. Kumar, P. Raghavan, S. Rajalopagan, A. Tomkins, 
Proceedings of the $9$th ACM Symposium on Principles of Database Systems, 
1 (1999).  
 
\bibitem{M67}  
S. Milgram,  
{\em Psych. Today} {\bf 2}, 60 (1967).  
 
\bibitem{N01}  
M.E.J. Newman,  
{\em Proceeding National Academy of Sciences} USA {\bf 98}, 404 (2001)  
 
\bibitem{ASBS00}  
L.A.N. Amaral, A. Scala, M. Barthelemy and H. E. Stanley, 
{\em Proceeding National Academy of Sciences} USA {\bf 97}, 11149 (2000)  

\bibitem{SM02}  
S. Maslov  and K. Sneppen,  
{\em Science}, {\bf 296}, 910 (2002) 

\bibitem{AB02}
R. Albert and A.-L. Barab\'asi 
{\em Reviews of Modern Physics} {\bf 74}, 47 (2002)

\bibitem{re3}
S.N. Dorogovtsev and J.F.F. Mendes, 
{\em Advances in Physics} {\bf 51}, 1079 (2002)

\bibitem{BB01}
G. Bianconi, A.-L. Barab\'asi 
{\em Europhysics Letters} {\bf 54}, 436 (2001).

\bibitem{CCDM02}
G. Caldarelli, A. Capocci, P. De Los Rios, M.A. Mu\~noz,
{\em  Physical Review Letters} 89, 258702 (2002). 

\bibitem{AJB99}  
R. Albert, H. Jeong, A.-L. Barab\'asi, 
{\em Nature} {\bf 401}, 130 (1999).  
  
\bibitem{g2}
K.-I Goh, B. Kahng and D. Kim, 
{\em Physical Review Letters}{\bf 88} 108701 (2002).

\bibitem{gin}
G. Bianconi, A. -L. Bar\'abasi,
{\em Europhysics Letters} {\bf 54}, 436 (2001).

\bibitem{re1}
P.L. Krapivsky, G.J. Rodgers and S. Redner,
{\em Physical Review Letters} {\bf 86}, 5401 (2001).

\bibitem{re2}
G. Ergun and G. J. Rodgers,
{\em Physica A} {\bf 303}, 261 (2002).

\bibitem{Ale}  
A. Vazquez, R. Pastor-Satorras, A. Vespignani 
{\em Physical Review E}  {\bf 65}, 066130 (2002)   

\bibitem{Ale2}  
R. Pastor-Satorras, A. Vazquez, A. Vespignani 
{\em Physical Review Letters} {\bf 87} 258701 (2001).  

\bibitem{mey}  
D. Meyer 
{\em Univ. of Oregon Route Views Archive Project} 
(http://archive.routeviews.org).  

\bibitem{N02}  
 M.E.J. Newman,  
{\em Physical Review Letters} {\bf 89}, 208701 (2002)  
 
\bibitem{WS98}  
D.J. Watts and S.H. Strogatz,  
{\em Nature} {\bf 393}, 440 (1998)  

\bibitem{Fre}  
L.C. Freeman
{\em Sociometry} {\bf 40} 35 (1977).  

\bibitem{Goh}  
K.-I. Goh, E.S. Oh, H. Jeong, D. Kim 
{\em Proceeding National Academy of Sciences} USA {\bf 99}, 12583 (2002).

\end{thebibliography}
\end{document}